\documentclass[aps, twocolumn,noshowpacs,preprintnumbers,amsmath,amssymb]{revtex4}

\usepackage{verbatim}
\usepackage{wasysym}

\usepackage{graphicx}
\usepackage{dcolumn}
\usepackage{bm}
\usepackage{wrapfig}

\begin{document}

\title{Sliding on a Nanotube: Interplay of Friction, Deformations and Structure}

\author{Hsiang-Chih Chiu}
\affiliation{School of physics, Georgia Institute of Technology, Atlanta, USA}
\author{Beate Ritz}
\affiliation{Institute of Physical Chemistry, University of Hamburg, Hamburg, Germany}
\author{Suenne Kim}
\affiliation{School of physics, Georgia Institute of Technology, Atlanta, USA }
\author{Erio Tosatti}
\affiliation{International School for Advanced Studies (SISSA), and CNR-IOM Democritos, Via Bonomea 265, 34136 Trieste, Italy; \\ International Centre for Theoretical Physics (ICTP), Strada Costiera 11, 34151 Trieste, Italy}
\author{Christian Klinke}
\affiliation{Institute of Physical Chemistry, University of Hamburg, Hamburg, Germany }
\author{Elisa Riedo}
\affiliation{ School of physics, Georgia Institute of Technology, Atlanta, USA }

\begin{abstract} 

The frictional properties of individual carbon nanotubes (CNTs) are studied by sliding an atomic force microscopy tip across and along its principle axis. This direction-dependent frictional behavior is found to correlate strongly with the presence of structural defects, surface chemistry, and CNT chirality. This study shows that it is experimentally possible to tune the frictional/adhesion properties of a CNT by controlling the CNT structure and surface chemistry, as well as use friction force to predict its structural and chemical properties.

\end{abstract}

\maketitle

The discovery of Carbon Nanotubes (CNT) has attracted considerable interests from both industrial and academic communities in the last two decades owing to their impressive physical properties \cite{1,2,3,4,5,6,7}. Applications in micro and nano-electro-mechanical systems such as actuators and sensors have been proposed, and some prototypes have already been demonstrated \cite{8,9,10}. Their exceptional mechanical properties and high aspect ratio also qualify them for mechanical reinforcement in composite materials \cite{11}. For most CNT applications, CNTs are in contact with their supporting surfaces, therefore it is imperative to understand their frictional properties \cite{12,13,14} and the influence of structural defects \cite{15}, and surface chemistry \cite{16,17} on these properties. Defects have been shown to significantly reduce the tensile and axial strengths of CNTs \cite{18,19,20,21,22} and to cause substantial increase in interlayer dissipation and friction when pulling concentric multiwalled CNTs \cite{23,24}. The static friction forces between the outer shell surfaces of two different multiwalled CNTs with dissimilar amount of structural defects have been measured by transmission electron microscopy (TEM), and larger friction forces have been found on the disordered CNT with rougher surface \cite{25}. Very recently, the frictional properties of individual CNTs have been studied with a nano-size atomic force microscope (AFM) tip sliding on a CNT lying on a substrate \cite{3}. A larger friction coefficient has been found when the tip is sliding perpendicular to the CNT axis, as compared with sliding along the tube axis. This behavior is explained by a deformation, like a lateral swaying (or a "hindered rolling") of the tube during the transverse sliding, which produces additional friction dissipation. This soft deformation mode is absent, or partially absent, when the tip slides along the CNT axis, thus, for the longitudinal sliding the friction force arises mostly only from sliding the hard nano-contact between the tip and the tube. The ratio between transverse and longitudinal friction per unit area is called friction anisotropy. Here, we show how structural defects, surface chemistry and possibly chirality can couple the transverse and longitudinal sliding, modulating nanotubes frictional properties. A simple analytical model has been developed to compute the amount of coupling, $\alpha$, between the transverse and longitudinal sliding, the "intrinsic" hard contact sliding shear strength, $\sigma^{int}$, and the soft "hindered rolling" shear strength, $\sigma^{HR}$. This model captures very well the observed experimental behavior indicating for all the CNTs a common supralinear decrease of $\alpha$ with increasing ratio between transverse and longitudinal friction. In the experiments, the amount of structural defects in the CNTs is controlled by using different growth methods, whereas the surface chemistry is controlled with after-growth chemical functionalization. To decouple the role of structural defects and molecular interactions at the surface, we have performed friction experiments on the same functionalized NT in different humidity, in this way humidity modulates the molecular interaction without changing the impact of structural defects. The model-experiment comparison shows that the coupling between transverse and longitudinal sliding, leading to small friction anisotropy values, is significantly increased by the presence of structural defects in the NTs. Isolated very high values of friction anisotropy, up to 14, for CNTs with no defects have been attributed to non-chiral CNTs. Furthermore, this investigation shows that the intrinsic friction and the transverse swaying all increase by increasing the amount of structural defects or increasing the strength of the molecular interaction between tip and tube. In fact, at higher relative humidity (R.H.), we find that the excess ambient moisture leads to larger friction forces in both tip sliding directions, in agreement with an enhancement of the strength of the intermolecular interaction between the carboxylic and hydroxyl groups on the CNTs and the silicon tip in presence of water molecules.

\begin{figure}[ht]
  \centering
  \includegraphics[width=0.45\textwidth]{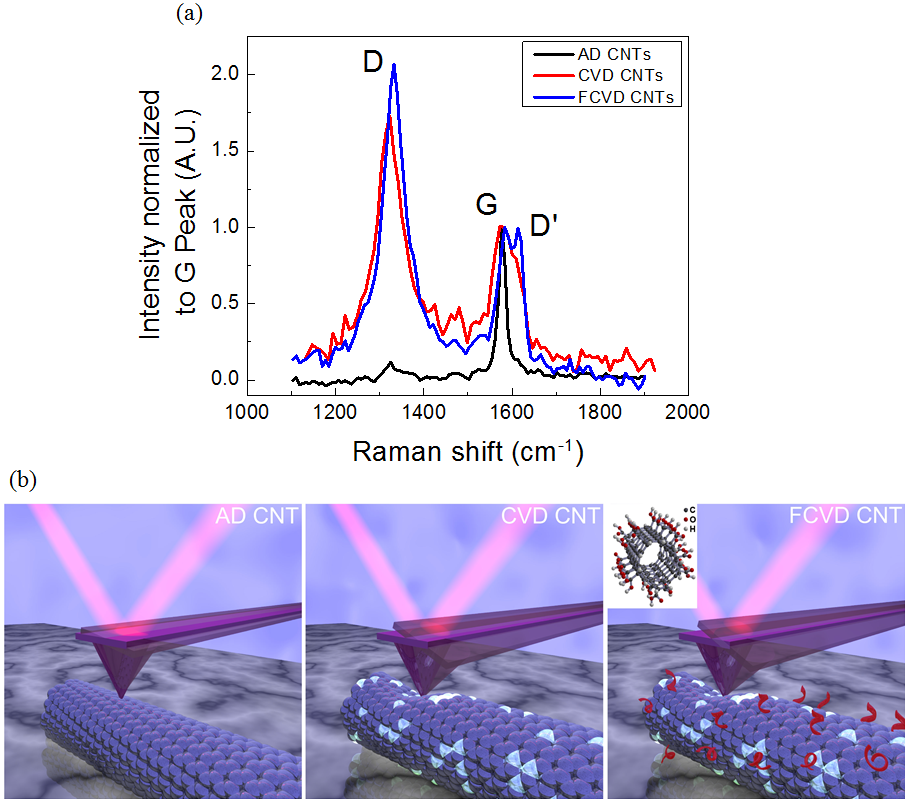}
  \caption{\textit{(a) Raman spectra of AD (Black), CVD (Red), and FCVD (Blue) CNTs. The $I_{D}/I_{G}$ ratio, as determined from Lorentzian lineshape fitting, is 0.34, 1.25 and 1.57 for AD, CVD and FCVD CNTs, respectively. An additional D' band in Raman spectra further indicates that FCVD CNTs are more disordered than CVD CNTs \cite{16}. (b) Cartoons show AD CNT, CVD and FCVD CNTs during longitudinal tip sliding. Due to the presence of defects and functional groups, there is more hindered rolling in CVD and FCVD CNTs, resulting in larger friction forces and cantilever deflections. The insertion shows a FCVD CNT functionalized with carboxylic (-COOH) groups on its external surface.}}
\end{figure}

In this study we report on measurements of the frictional properties of supported multiwalled CNTs produced by Arc Discharge (AD), and Chemical Vapor Deposition (CVD), as well as chemically functionalized CVD (FCVD) CNTs. The chemical functionalization is obtained by treatment with nitric acid, resulting in carboxylic (-COOH) and hydroxyl (-COH) groups bonded to the CNT surface. It is well known that CVD grown CNTs possess more structural defects than the AD grown ones. TEM images and Raman spectra (Fig. 1) show that CVD CNTs have abundant defects. Furthermore, functionalizing CVD CNTs with nitric acid results in additional defects \cite{16}. Figure 1(a) shows the Raman spectra of the different types of CNTs studied here. Significant increase in the Raman D to G-band intensity ratio, $I_{D}/I_{G}$, indicates that CVD and FCVD CNTs are highly disordered \cite{26}. The friction forces and the topography are acquired simultaneously by AFM for each individual CNT laying on a silicon substrate. A nano-size AFM tip is used to slide on top of a CNT perpendicularly (transverse sliding, T) and in parallel (Longitudinal sliding, L, see Figure 1(b)) to the tube axis to acquire the friction forces, $F_{F}$, and the T-L friction anisotropy. After an appropriate CNT is found, the sample is rotated until the CNT axis is perpendicular to the AFM cantilever axis, i.e., parallel to the fast scan direction. In this configuration, the longitudinal friction measurements are performed. The sample is subsequently rotated by 90$^{\circ}$ for the transverse friction measurements on the same part of the CNT. This ensures similar contributions to the friction forces from structural defects and eventually chirality. 

Figure 2 (a) and (b) show typical AD and CVD CNT images before and after the 90$^{\circ}$ sample rotation. In Figure 2(c) and (d), the corresponding transverse friction forces $F^{T}_{F}$ and longitudinal friction forces $F^{L}_{F}$ as a function of normal loads, $F_{N}$, are shown. To take into account the different tip's and CNT's radii, $R_{Tip}$ and $R_{NT}$, the friction forces are normalized to $(1/R_{Tip} + 1/2 R_{NT})^{2/3}$ \cite{4}. As consistently observed in our experiments, it is evident that the friction coefficient is the largest for FCVD CNTs and the smallest for AD CNTs in both scanning directions. Furthermore, the normalized longitudinal friction forces $F^{L}_{F}$ on CVD and FCVD CNTs are clearly more than one order of magnitude larger than the force on AD CNTs. These results indicate that structural defects increase the dissipation and thus the frictional forces between the CNT and the AFM tip during sliding. Furthermore, the -O(OH) and -OH functional groups present on the FCVD NTs play a crucial role in increasing the strength of the molecular interaction between the silicon/silicon oxide tip and the NT, therefore increasing the friction force.

In the macroscopic world, the friction force between two objects is generally linearly dependent on the normal force that compresses them together and is independent of their apparent contact area. However, at the nanoscale, this linear relationship is no longer valid and one has to consider the single contact geometry. The friction force will thus be proportional to their contact area $A$, which is a function of the shape, size and elasticity of two nanoscale objects. In the case of an AFM tip sliding on a CNT, the tip-CNT contact area $A$ can be calculated with the modified Hertz theory \cite{27}. The friction force can be described by

\begin{equation} 
F_{F} = \sigma \cdot A \cdot (F_{N} + F_{Adh}) = \sigma \cdot \gamma \cdot (F_{N} + F_{Adh})^{2/3}
\end{equation}

where $\sigma$ and $F_{Adh}$ are respectively the shear strength and adhesion force between the tip and the CNT \cite{4}. The constant $\gamma$ contains information about $R_{NT}$, $R_{Tip}$ and the Young's modulus of both surfaces. The derivation of equation (1) is described in literature \cite{27,28,29}. To further compare the frictional behavior of CNTs with different structural disorder, the data shown in Figures 2(c) and (d) are fitted with equation (1) to obtain $\sigma$ and $F_{Adh}$ from each CNT in the transverse and longitudinal directions ($\sigma_{T}$ and $\sigma_{L}$, respectively). All fitted shear strengths $\sigma_{T,L}$ as a function of $R_{NT}$ are shown in Figure 3(a). For AD CNTs, the transverse shear strength $\sigma^{AD}_{T}$ has values between 0.04 $\pm$ 0.01 GPa and 0.28 $\pm$ 0.005 GPa. The transverse shear strengths for CVD and FCVD CNTs, $\sigma^{CVD}_{T}$ and $\sigma^{FCVD}_{T}$, are much larger than the ones in AD CNT. The value of $\sigma^{CVD}_{T}$ fluctuates between 0.24 $\pm$ 0.02 and 0.38 $\pm$ 0.01 GPa while $\sigma^{FCVD}_{T}$ can be as large as 0.77 $\pm$ 0.12 GPa. Even more significant difference exists in $\sigma_{L}$. The longitudinal shear strength for AD CNTs, $\sigma^{AD}_{L}$, is independent of $R_{NT}$ between 6 and 19 nm and remains around 0.02 GPa. But the longitudinal shear strengths for CVD and FCVD CNTs, $\sigma^{CVD}_{L}$ and $\sigma^{FCVD}_{L}$, can be as much as 10 to 30 times larger than  $\sigma^{AD}_{L}$. 

\begin{figure}[ht]
  \centering
  \includegraphics[width=0.45\textwidth]{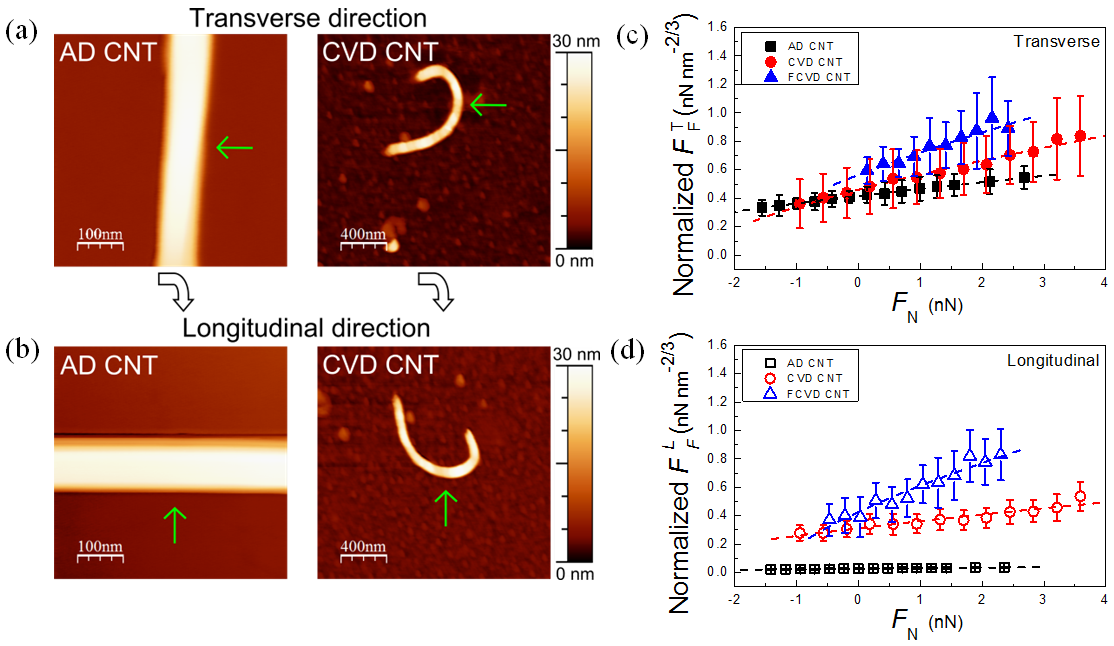}
  \caption{\textit{(a) and (b) The AFM images of CNTs before and after 90$^{\circ}$ sample rotation. The $R_{NT}$ is 15.1 $\pm$ 0.7 nm and 11.4 $\pm$ 0.4 nm for AD and CVD CNTs, respectively. Green arrows indicate the locations where the friction measurements take place; (c) and (d) Normalized friction forces $F_{F}$ vs. normal loads $F_{N}$ for AD CNTs (Black), CVD CNTs (Red) and FCVD CNTs (Blue) in the transverse (filled symbol) and longitudinal (open symbol) sliding directions. The dashed lines are obtained by the fitting friction force data to equation (1). For AD CNT, the fitted $\sigma^{AD}_{T}$ and $\sigma^{AD}_{L}$ are 0.22 $\pm$ 0.002 GPa and 0.016 $\pm$ 0.001 GPa, respectively, representing a friction anisotropy of 13.7. For the CVD CNT, $\sigma^{CVD}_{T}$ and $\sigma^{CVD}_{L}$ are 0.38 $\pm$ 0.01 GPa and 0.20 $\pm$ 0.01 GPa, respectively, corresponding to a friction anisotropy of only 1.9. For functionalized CNTs, the fitted $\sigma^{FCVD}_{T}$ and $\sigma^{FCVD}_{L}$ are both 0.61 $\pm$ 0.03 GPa with a friction anisotropy of only 1.}}
\end{figure}

For an ideal, non-chiral CNT, the friction anisotropy, here defined as $a \equiv \sigma_{T}/\sigma_{L}$, originates from the effect of lateral swaying also called "hindered rolling" during the transverse sliding, which opens a new channel for energy dissipation. On the contrary, during the longitudinal sliding, this extra energy dissipation is partially absent, giving rise to smaller friction forces which are mostly due to the "intrinsic" force required for sliding the hard contact between the tip and the nanotube. The experimental results reported in Figure 3(a) can be explained by considering the presence of structural defects, chemical functional groups and CNT chirality, which all contribute to the intrinsic friction, the swaying deformations, and the coupling between the transverse and the longitudinal motion of the tube during tip sliding so that a partial lateral swaying will be present also during the longitudinal sliding. This coupling will then possibly increase the longitudinal friction force and decrease the transverse one. A molecular dynamics simulation has indeed shown that the friction anisotropy of a non-chiral CNT can be as large as 20 but drops to only 2 for a chiral CNT, due to a screw-like motion of the tip during the longitudinal sliding \cite{3}. Figure 3(b) shows the ratio between $\sigma_{T}$ and $\sigma_{L}$ for all the measured CNTs as a function of $R_{NT}$. For AD CNTs, the largest friction anisotropy is 13.7 and the smallest is 2.5. Based on the above discussion, and the low amount of defects present in AD CNTs, we argue that in Figure 3(b) the data points shown in the shaded area correspond to chiral and more asymmetric AD CNTs, while the ones displayed outside that area correspond to non-chiral or more symmetric AD CNTs. For CVD CNTs, which are rich of structural defects, besides the chirality, defects can cause a strong coupling between the transverse and the longitudinal motion during the tip sliding, i.e. small anisotropy. The friction anisotropy of CVD CNT lies indeed within a range between 1.5 and 5.7, always inside the shaded area of Figure 3(b). For FCVD CNTs, which present more structural defects and a stronger interaction with the AFM tip due to the presence of reactive functional groups on their surface, the friction anisotropy is always below 2.

\begin{figure}[ht!]
  \centering
  \includegraphics[width=0.38\textwidth]{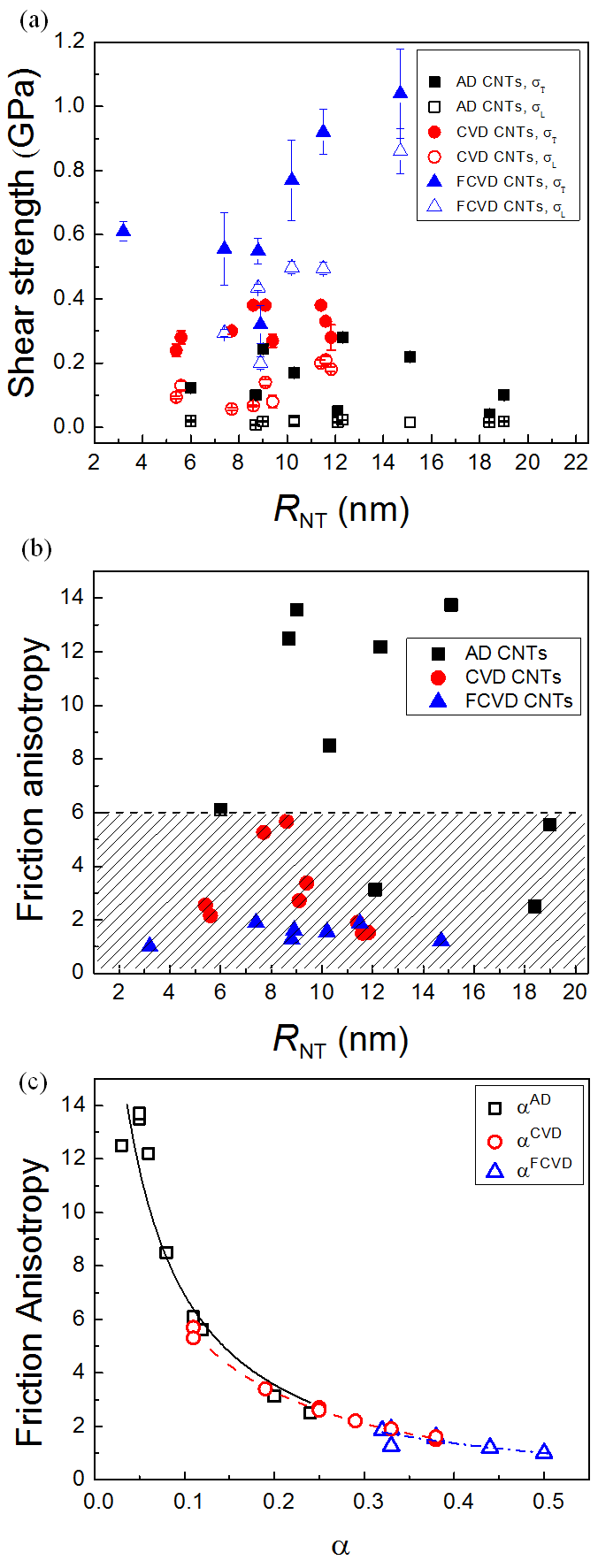}
  \caption{\textit{(a) Shear strengths obtained from fitting the friction force data to equation (1) as a function of $R_{NT}$. The error bars are determined from the fitting procedure. (b) Friction anisotropy of all investigated CNTs as a function of $R_{NT}$. The shaded area indicates strong defect and chirality induced coupling of tip motion during sliding. (c) Friction anisotropy as a function of the coupling parameter $\alpha$ obtained from the model. The lines are the best fits of each data to the equation $\sigma_{T}/\sigma_{L} = [ 1 + ( 1 - \alpha ) \cdot P ] / ( 1 + \alpha \cdot P )$, where $P = \sigma^{HR}/\sigma^{int}$, derived from equation (2) and (3). When $\alpha = 0$, the maximum friction anisotropy will be $P + 1$. See Table I for fitting results.}}
\end{figure}

To gain a fundamental understanding of the physical and chemical processes giving rise to the frictional behavior presented in the Figures 2 and 3, we have developed a simple model with which we extract quantitative information about the amount of transverse-longitudinal coupling, as well as the magnitude of the "intrinsic" and "hindered rolling" shear strengths.  The constitutive equations of the model are the following:

\begin{equation} 
\sigma_{L} = \sigma^{int} + \alpha \cdot \sigma^{HR}
\end{equation}

\begin{equation} 
\sigma_{T} = \sigma^{int} + (1 - \alpha) \cdot \sigma^{HR}
\end{equation}

\begin{table*}[ht!]
\caption{The maximum and minimum friction anisotropy, average coupling parameter,        intrinsic strength, "hindered rolling" shear strength and the fitted $P$ from  the model for three different CNT systems.}
    \begin{tabular}{|c|c|c|c|c|c|c|} 
				\hline
        \textbf{CNTs}			& $a^{Max}$	& $a^{Min}$ & $< \alpha >$ & $< \sigma^{int} >$ / GPa & $< \sigma^{HR} >$ / GPa & $P$ \\ \hline
        \textbf{AD CNT}  	& 13.7	& 2.5	& 0.10 $\pm$ 0.02	& 0.005 						& 0.15 $\pm$ 0.03	& 28.7 $\pm$ 4.3  \\ \hline
        \textbf{CVD CNT}	& 5.7		& 1.5	& 0.25 $\pm$ 0.03	& 0.022 $\pm$ 0.003	& 0.40 $\pm$ 0.03	& 15.8 $\pm$ 0.7	 \\ \hline
        \textbf{FCVD CNT} & 1.9		& 1.0	& 0.32 $\pm$ 0.07	& 0.08 $\pm$ 0.07		& 0.75 $\pm$ 0.13	& 5.5  $\pm$ 3.6	 \\
    \hline
    \end{tabular}
\end{table*}

\begin{table*}[ht!]
\caption{The measured shear strengths $\sigma_{T,L}$, the friction anisotropy $a$ and the obtained $\sigma^{int,HR}$ of two FCVD CNTs at different relative humidity (R.H.).}
	
    \begin{tabular}{|p{10mm}|p{16mm}|p{68.6mm}|p{68.6mm}|} 
				\hline
         &  & FCVD CNT \# 1 ($R_{NT}$ = 14.7 $\pm$ 0.1 nm) & FCVD CNT \# 2 ($R_{NT}$ = 11.5 $\pm$ 0.5 nm) 
    \end{tabular}

    \begin{tabular}{|p{10mm}|p{16mm}|p{16mm}|p{16mm}|p{16mm}|p{16mm}|p{16mm}|p{16mm}|p{16mm}|p{16mm}|} 
         \textbf{R.H.} & $\sigma^{int}$ / GPa & $\sigma_{T}$ / GPa & $\sigma_{L}$ / GPa & $\sigma^{HR}$ / GPa & $a$ & $\sigma_{T}$ / GPa & $\sigma_{L}$ / GPa & $\sigma^{HR}$ / GPa & $a$ 
    \end{tabular}

    \begin{tabular}{|p{10mm}|p{16mm}|p{16mm}|p{16mm}|p{16mm}|p{16mm}|p{16mm}|p{16mm}|p{16mm}|p{16mm}|} 
				\hline
        \textbf{20\%}	& 0.1 $\pm$ 0.1	& 0.37 $\pm$ 0.04	& 0.38 $\pm$ 0.04	& 0.52 $\pm$ 0.10	& 1.0 $\pm$ 0.1	& 0.39 $\pm$ 0.07 & 0.33 $\pm$ 0.02 & 0.50 $\pm$ 0.01 & 1.2 $\pm$ 0.1 \\ \hline
        \textbf{36\%}	& 0.2	$\pm$ 0.2	& 1.0 $\pm$ 0.1		& 0.86 $\pm$ 0.07	& 1.50 $\pm$ 0.14	& 1.2 $\pm$ 0.2	& 0.92 $\pm$ 0.07	& 0.49 $\pm$ 0.02 & 1.18 $\pm$ 0.01 & 1.8 $\pm$ 0.1 \\ 
    \hline
    \end{tabular}

\end{table*}

These equations are then solved for each set of experiments reported in Figure 3(a). $\sigma^{HR}$ depends on the size of the NT, presence of defects, chirality, and molecular interaction/adhesion between tip-NT and NT-substrate. The parameter $\alpha$ represents the coupling between the transverse and the longitudinal motion of the tube during tip sliding due, for example, to structural defects, and CNT chirality. In this model $\alpha = 0$ for zero coupling and $\alpha = 1/2$ for maximum coupling. This coupling causes larger longitudinal friction forces and suppresses the effect of hindered rolling during transverse sliding. For AD CNTs with no or minimum defects, the intrinsic shear strength $\sigma^{int-AD}$ can be considered to be the one between an AFM tip and HOPG, which was reported to be 5 MPa \cite{12,28}. Hence from equation (2) and (3), knowing $\sigma^{T}$ and $\sigma^{L}$ from the experiments, we obtain $\sigma^{HR-AD}$ which varies between 0.046 GPa and 0.293 GPa, and $\alpha^{AD}$, which varies between 0.03 and 0.24 for all studied AD CNTs. The values of $\alpha^{AD}$ obtained for different CNTs are found to collapse onto a single curve as a function of the friction anisotropy, as shown in Figure 3(c), where $\alpha$ decreases with increasing friction anisotropy, indicating that the model well captures the physics of this phenomenon. In this plot it is possible to observe some extremely small values of $\alpha$. We argue that these values can be ascribed to non-chiral AD CNTs. Variations of $\alpha$ and $\sigma^{HR-AD}$ are instead probably due to variations in the external tube chirality and/or presence of defects. The value of $\sigma^{HR-AD}$ can be 20 to 60 times larger than $\sigma^{int-AD}$; therefore the effect of hindered rolling is the main contribution to the measured $\sigma^{AD}_{T}$ . For CVD CNTs, which have more defects, $\sigma^{int-CVD}$ is unknown but can be estimated in the following way. Figure 3(c) indicates that it is reasonable to assume that CNTs with the same friction anisotropy shall have the same coupling parameter $\alpha$. Thus by comparing AD CNTs and CVD CNTs with the same friction anisotropy, we can assume $\alpha^{CVD} = \alpha^{AD}$ and estimate the value of $\sigma^{int-CVD}$ from the experimental data of $\sigma^{CVD}_{T}$ and $\sigma^{CVD}_{L}$. As a result, $\sigma^{int-CVD}$ is found to be 0.022 GPa, which is 4 times larger than $\sigma^{int-AD}$. This increased  $\sigma^{int-CVD}$ can be related to the presence of defects which are known to increase the shear strength during telescopic rotation of concentric CNTs \cite{23,24}. Furthermore, it is found that $\sigma^{HR-CVD}$ varies between 0.29 GPa and 0.54 GPa, values up to one order of magnitude larger than  $\sigma^{HR-AD}$. Next we obtain that $\alpha^{CVD}$ varies in the range between 0.11 and 0.38, values much larger than $\alpha^{AD}$, as presented in Figure 3(c). This signifies that the coupling effect in CVD CNTs during tip sliding is indeed stronger than in AD CNTs owing to the structural disorder, which might also obscure any effect from CNT chirality. There is a concern about the CNT curvature effect on $\sigma^{HR-CVD}$ since during the transverse sliding the bent parts of CNT next to the tip-CNT contact might serve as anchors that further suppress CNT rolling on the surface. However, no relation between $\sigma^{HR-CVD}$ and bending radius $R_{bending}$ of CNT is found within current results. For FCVD CNTs presenting surfaces functionalized with carboxylic groups, $\sigma^{int-FCVD}$ is found to be 0.08 $\pm$ 0.07 GPa, which is 4 and 16 times larger than $\sigma^{int-CVD}$ and $\sigma^{int-AD}$, respectively. Furthermore, $\sigma^{HR-FCVD}$ is between 0.5 and 1.1 GPa, significantly larger than the previous types of CNTs. All these results clearly indicate that both the intrinsic friction and the "hindered rolling" dissipation are larger in nanotubes rich of structural defects and with a stronger interaction with the sliding silicon tip. For example, for FCVD CNTs, a chemical bond between the tip and the FCVD CNT due to the presence of functional groups might form at the onset of the tip-CNT contact. At the subsequent stage, this bonding will break when tip slides on the CNT. The forming and breaking of chemical bonds result in larger $\sigma^{int}$ for the FCVD CNT compared to the other two types of CNT systems. The pulling, stretching and breaking of chemical bonds will produce additional friction dissipation channels and thus larger $\sigma^{HR-FCVD}$. Table I summarizes the aforementioned parameters of these three types of CNT systems.

To understand the role of molecular interactions at the tip-CNT interface, we have performed friction experiments on FCVD CNT by keeping unaltered humidities, in this way humidity can change the molecular interaction between the tip and the CNT. The equations-experiment comparison shows that the intrinsic friction and the transverse swaying increase by the increased strength of the molecular interaction between tip and tube. In fact, at higher relative humidity (R.H.), from 20\% to 36\%, we find that the excess ambient moisture leads to larger friction forces in both tip sliding directions. This increase indicates an enhancement of the strength of the molecular interaction between the CNT and functional groups and the silicon tip in presence of water molecules. This increased strength is possibly due to water mediated deprotonation of the carboxylic groups and formations of charged and very reactive functional groups on the nanotubes. The measured shear strengths and friction anisotropies at different R.H. are reported in Table II. At larger R.H.=36\%, the obtained shear strengths are increased in both sliding directions. Similar analysis with the proposed model also shows that $\sigma^{int}$ and $\sigma^{HR}$ both increase from 0.1 $\pm$ 0.1 GPa to 0.2 $\pm$ 0.18 GPa and 0.5 $\pm$ 0.1 GPa to 1.5 $\pm$ 0.2 GPa, respectively. On the other side, the friction anisotropy only increases mildly from an average value of $a \approx 1$ to $a \approx 1.5$.

In conclusion, we have measured and modeled the friction forces between an AFM tip and individual CNTs with different structural disorder, different surface chemistry and in different humidity environments. By comparing the friction forces in different CNT systems, and environments we conclude that both the intrinsic friction and the "hindered rolling" dissipation are larger in nanotubes rich of structural defects and with a stronger interaction with the sliding silicon tip. Compared to defect-free AD CNTs, the maximum intrinsic and "hindered rolling" shear strengths are enhanced by 4 and 2.6 times respectively for defective CVD CNTs, 16 and 5 times for chemically functionalized FCVD CNTs. Furthermore, structural defects and chirality can cause strong coupling between the transverse and longitudinal dissipation modes during tip sliding on CNTs. The friction anisotropy, defined by the ratio between transverse and longitudinal shear strengths is found to decrease in the presence of defects. The maximum friction anisotropy of 13.7 is found in defect-free, possibly non-chiral, AD CNTs. This anisotropy becomes less than 6 for CVD CNTs with defects and further reduced to less than 2 for FCVD CNTs with more defects and surface functionalization. Understanding the frictional property of CNTs is important for CNT applications such as CNT reinforced composite materials. For example, employment of CNTs with structural defects is found to increase the fracture toughness of ceramic composite materials owing to the enhanced interfacial friction force between CNTs and neighboring ceramic materials \cite{30}. Our finding provides a better understanding of the tribological properties of individual carbon nanotube at the nanoscale. More generally, it might help with the development of stronger composite materials and better strategies for nano-object manipulation on surfaces.

\section*{Experimentals}

The AD and CVD CNTs are purchased from n-Tec (Norway) and Bayer Material Science (Germany), respectively. The FCVD CNTs are functionalized mainly with the carboxylic groups following the protocol described in the supporting information and Ref. \cite{16}. All CNTs are deposited on an acetone cleaned silicon substrate for friction measurements performed with a Veeco Nanoscope IV Multimode AFM at room temperature and in ambient environment. Details about Raman measurements, sample preparation and TEM data regarding CNT structure, size and number of walls are reported in the supporting information. A cantilever (PPP-LFMR, Nano and More) with a typical spring constant $k_{N} = 0.2$ N m$^{-1}$ is used. The spring constant of cantilever is calibrated with the Sader's method \cite{31}. The lateral sensitivity is calibrated by the wedge method \cite{32}. The $R_{NT}$ and average $R_{Tip} = 50$ nm are inferred directly from the AFM images of CNTs. The friction forces are measured usually beginning with $F_{N} \approx 3$ nN and decreased stepwise until the tip is out of contact with the sample. The tip velocity is kept at 1 $\mu$m $s^{-1}$ for all measurements. All measurements discussed in Fig. 1, 2, and 3 are performed at R.H. of about 40\%. In Table II, we repeat the measurements on two FCVD CNTs at controlled R.H. of 20\% and 36\%. The fluctuation of R.H. is less than 1\% during the entire experiment.

\section*{Acknowledgements}

H.-C.C. and S.K. were financially supported by the Office of Basic Energy Sciences of the DOE (DE-FG02-06ER46293). 
E.R. acknowledges the DOE (DE-FG02-06ER46293) and NSF (DMR-0120967 and DMR-0706031) for summer salary support. 
E.T. acknowledges discussions with X.H. Zhang, and support by CNR Project FANAS/AFRI, and by PRIN Contract 20087NX9Y7.


\end{document}